\documentclass[twocolumn,journal]{IEEEtran}
\pdfoutput=1 
\usepackage{graphicx}
\usepackage[english]{babel}
\usepackage{amssymb}
\usepackage{amsbsy}
\usepackage{multirow}
\usepackage{slashbox}
\usepackage{algorithmic}
\usepackage{algorithm}

\title{A Flexible LDPC code decoder with a Network on Chip as underlying interconnect architecture}

\author{Carlo Condo, Guido Masera, \emph{Senior Member IEEE} }

\begin{document}

\maketitle

\begin{abstract}
LDPC (Low Density Parity Check) codes are among the most powerful and widely adopted modern error correcting codes. 
The iterative decoding algorithms required for these codes involve high computational complexity and 
high processing throughput is achieved by allocating a sufficient number of processing elements (PEs). 
Supporting multiple heterogeneous LDPC codes
on a parallel decoder poses serious problems in the design of the interconnect structure for such PEs.
The aim of this work is to explore the feasibility of NoC (Network on Chip) based decoders, where full flexibility 
in terms of supported LDPC codes is obtained resorting to an NoC to connect PEs.
NoC based LDPC decoders
have been previously considered unfeasible because of the cost overhead associated to packet management
and routing. On the contrary, the designed NoC adopts a  
low complexity routing, which introduces a very limited cost overhead with respect to architectures 
dedicated to specific classes of codes. Moreover the paper proposes an efficient configuration technique,
which allows for fast on--the--fly switching among different codes.
The decoder architecture is scalable and VLSI synthesis results are presented for several
cases of study, including the whole set of WiMAX LDPC codes, WiFi codes and DVB-S2 standard.
\end{abstract}


\section{Introduction}
\label{intro}

The original introduction of LDPC (Low Density Parity Check) codes \cite{gallager} 
and their more recent rediscovery by MacKay and Neal \cite{mackay} stimulated a large amount of studies on both decoding algorithms and hardware implementations.
LDPC codes are included in a growing number of applications such as IEEE 802.11n \cite{wifi}, IEEE 802.16e \cite{wimax} and DVB-S2 \cite{dvbs2}.
Therefore flexible decoders capable of working for multiple codes are receiving a significant attention.


Flexibility in terms of supported codes and executed decoding algorithms can be obtained resorting to either parameterized processing elements (PE) or specialized programmable processors. Both solutions have been proved to provide enough flexibility at the processing level \cite{vacca2007} \cite{flexichap}. In order to obtain the same flexibility at the level of inter-PE communication, proper interconnect structures must be adopted, capable of supporting the different communication needs that are specific of each code.
Dedicated interconnect structures, with excellent characteristics of efficiency have been proposed for single codes or classes of codes 
(see for example \cite{Liu_2009} and \cite{Chen_2010}). In this kind of approach, the specific inter-processor communication needs are mapped onto low-cost interconnect structures. A relevant example is given by the class of quasi-cyclic LDPC codes \cite{fossorier}, where the parity check matrix (${\bf H}$) is structured as a set of sub-matrices that can be considered as circular shifted versions of the identity matrix. This particular form of ${\bf H}$ allows for relatively simple interconnect structures composed by barrel-shifters. 
Clearly the same approach cannot be adopted in the case of a fully flexible decoder, which has to support heterogeneous ${\bf H}$ matrices, with no common characteristics. In this case, the interconnect structure can be designed as an Application Specific NoC (ASNoC) \cite{benini}, that is an NoC  carefully tailored to the specific application to be supported. Contrariwise to the more common case of NoCs designed to connect heterogeneous processing tasks or Intellectual Property (IP) units \cite{benini_demicheli} \cite{goossens} ({\it Inter-IP} NoCs), in this work 
a kind of {\it Intra-IP} NoC is proposed to interconnect in a flexible way multiple homogeneous PEs that concurrently implement a channel decoding IP.\\
An NoC based flexible decoder includes a set of nodes, each one associated to a local PE and directly connected to a small subset of other nodes in the network. The required connectivity is obtained by means of routers, which decide the path for each data to be sent from a source node to a destination. 
Given experimental results show that: (i) the proposed fully flexible NoC based decoder achieves  
throughput values compliant with several standards; (ii) area overhead introduce by the NoC interconnect 
architecture is limited; (iii) on--the--fly reconfiguration of the NoC based decoder is feasible
to switch between different codes at no additional latency. 

In this paper, Section \ref{sec:LDPCdecode} summarizes the adopted decoding algorithm, while Section 
\ref{sec:NoCdecode} describes the NoC approach to LDPC code decoding. Section \ref{sec:architecture}
details the architecture of the single processing element and Section \ref{sec:configflow} explains the steps necessary to configure the decoder, while Section \ref{sec:results} provides 
results on the designed decoders in terms of achievable throughput, occupied area and comparisons with
other implementations. Conclusions are drawn in Section \ref{sec:conclusions}. 

\section{LDPC decoding}
\label{sec:LDPCdecode}

An LDPC code is a linear block code characterized by a very sparse ${\bf H}$ matrix. 
Columns (index $j$) of ${\bf H}$ are associated to received bits, while rows (index $m$) correspond to 
parity check constraints. In the layered decoding method \cite{Guilloud2007}, parity check constraints are grouped 
in layers and each layer is associated to a component code. Layers are decoded in sequence
by propagating extrinsic probability values from one layer to the following one \cite{hocevar_lay}. When
all layers have been decoded, one iteration is complete and the overall process can be 
iteratively repeated up to the desired level of reliability.
Layered decoding is known to approximately provide a factor two speed--up in terms of convergence speed
over the two--phase decoding method \cite{Guilloud2007}.

The layered decoding algorithm is now briefly reviewed
following the notation adopted in \cite{hocevar_lay}. $L(c)$
indicates the logarithmic likelihood ratio (LLR) of symbol $c$ ($L(c)=log(P\left\{c=0\right\}/P\left\{c=1\right\})$).
According to this notation, for each ${\bf H}$ column $j$, bit LLR $L(q_j)$ is initially set to the 
corresponding received soft value. Then, for all parity constraints $m$ in a given layer, the following
operations are executed:
\begin{equation}
	\label{lq}
      L(q_{mj}) =  L(q_{j}^{\rm (old)}) - R_{mj}^{\rm (old)}
\end{equation}
\begin{equation}
	\label{amj}
     A_{mj} =  \sum_{n\in N(m), n\neq j} \Psi ( L(q_{mn}))
\end{equation}
\begin{equation}
	\label{smj}
	 s_{mj} =  \prod_{n\in N(m), n\neq j} Sign ( L(q_{mn}))
\end{equation}
\begin{equation}
\label{lq2}
      R_{mj}^{\rm (new)} = -s_{mj} \Psi(A_{mj})
\end{equation}
\begin{equation}
\label{lq3}	
      L(q_{j}^{\rm (new)}) =  L(q_{mj}) + R_{mj}^{\rm (new)}
\end{equation}
$L(q_{j}^{\rm (old)})$ is the extrinsic information received from the previous layer and updated in
(\ref{lq3}) to be propagated to the succeeding layers.
Term $R_{mj}^{\rm (old)}$ pertaining to element $(m,j)$ of ${\bf H}$ is used to compute equation (\ref{lq}); 
the same amount is then updated in (\ref{lq2}), $R_{mj}^{\rm (new)}$, and stored to be used again in the following iteration.
In (\ref{amj}) and (\ref{smj}), $N(m)$ is the set of all bit indexes that are
connected to parity constraint $m$.  
Finally, $\Psi(\cdot)$ is a non--linear non--limited function usually replaced with a simpler
approximation. In this work, the normalized min--sum approximation \cite{fossorier_mihaljevic} is used
leading to the following formulation of (\ref{amj}):
\begin{equation}
	\label{amj1}
     A_{mj}^1 = min_{n\in N(m)}(|L(q_{mn})|)
\end{equation}
\begin{equation}
	\label{amj2}
     A_{mj}^2 = min_{n\in N(m),n\neq t}(|L(q_{mn})|)
\end{equation}
\noindent where $t$ is the index related to first minimum $A_{mj}^1$, while $A_{mj}^2$ is the second minimum.  
Equation (\ref{lq2}) is also changed to
\begin{equation}
\label{rmj1}
      R_{mj}^{\rm (new)} = \left\{
\begin{array}{ll}
      -s_{mj} \cdot A_{mj}^1/\alpha & {\rm when} \,\, |L(q_{mj})|\neq A_{mj}^1 \\
      -s_{mj} \cdot A_{mj}^2/\alpha & {\rm otherwise} \\
\end{array}
\right.
\end{equation}
\noindent where $\alpha$ is a normalization factor $\geq1$ that reduces performance degradation  
due to the min--sum approximation \cite{fossorier_mihaljevic}.

\section{NoC based decoding}
\label{sec:NoCdecode}

\begin{figure}[t]
	\centering
		\vspace{-60pt}
		\hspace{-110pt}
		\includegraphics[scale=0.5]{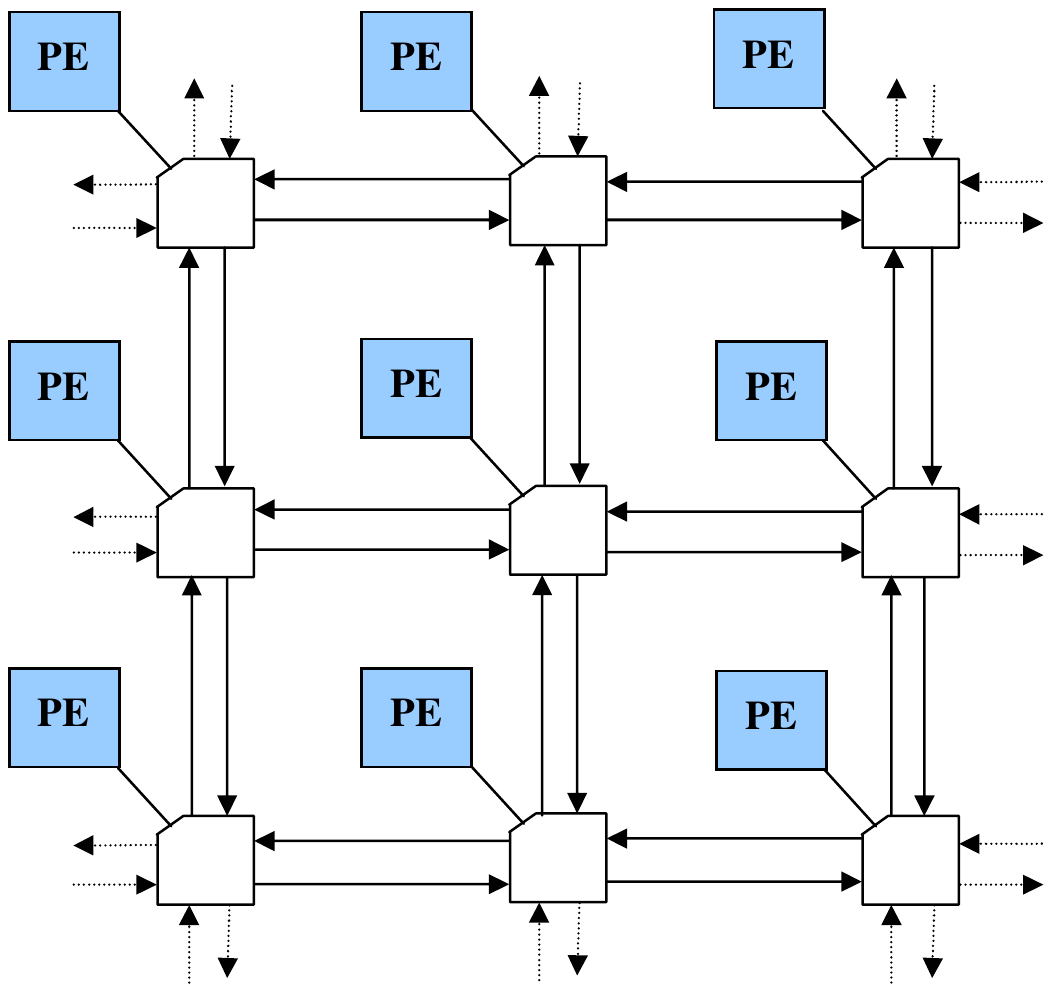}
		\hspace{-120pt}
		\vspace{-220pt}
	\centering
	\caption{NoC torus mesh topology}
			\vspace{-5pt}
	 \label{fig:torus}
\end{figure}

Partially parallel decoding architectures
are implemented by allocating a number of concurrent PEs, each one executing 
equations (\ref{lq}) to (\ref{rmj1}) on different sets of parity check constraints.
A proper interconnect structure must be used to deliver extrinsic information from one 
processor to another. Efficient dedicated networks have been proposed to 
provide inter--processor communication  in the case of specific families of LDPC codes.
This work focuses on complete flexibility of the decoder and therefore 
no assumption is made on the structure of LDPC codes to be supported.
To achieve such a large flexibility, the possible use of NoC based interconnect architectures has already been 
suggested and partially explored in \cite{theocharides} and \cite{masera_noc}: however, a complete evaluation
of the potential of the NoC--based approach in terms of achievable performance and implementation complexity
is not available. 

\begin{figure}[t]
	\centering
		\vspace{-90pt}
		\hspace{-120pt}
		\includegraphics[scale=0.6]{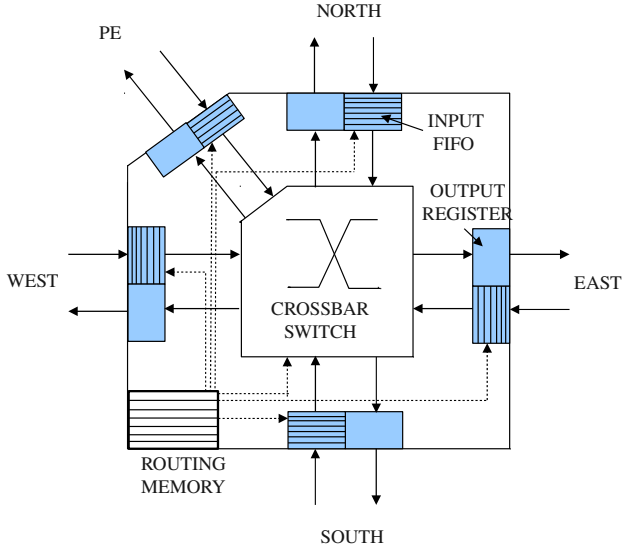}
		\hspace{-115pt}
		\vspace{-185pt}
	\centering
	\caption{NoC routing element}
		\vspace{-5pt}
	 \label{fig:router}
\end{figure}

The studied NoC-based decoder architecture relies on a 2D torus mesh topology (Figure \ref{fig:torus}), 
where each node has five input--output ports: four ports are connected to neighboring nodes, while the fifth port
connects to the local PE (Fig. \ref{fig:router}), which includes processing and memory
components required to execute the assigned decoding tasks. 
A simple input queuing architecture is adopted for the node and therefore
each input port has a first-in first-out memory queue (FIFO). A crossbar connects these FIFOs to output registers,
which are directly attached to output ports. Such simple and regular structure is well suited 
for VLSI design. The number of PEs in the NoC is much lower than the number of parity constraints 
in the ${\bf H}$ matrix.
Therefore, in a full decoding iteration, each PE sequentially serves multiple parity check constraints, according to a
defined scheduling: the lack of data dependencies in a layer implies that the parity check constraints
belonging to a given layer can be served at the same time by concurrent PEs.

In the straightforward approach to NoC based decoding, RPs deliver messages containing 
three elements: a payload that carries the extrinsic information, a header containing the 
identifier of the destination node and used for routing purposes, and the identifier of the 
parity check constraint mapped to the destination node. This kind of organization introduces a
relevant implementation overhead: first, identifiers associated to destination nodes  
tend to increase packet length and input FIFO size; second, 
a routing algorithm must be run at nodes to decide on the proper path for incoming packets
and to control accordingly crossbar and FIFOs.

However, the characteristics of the supported application can be exploited to eliminate this implementation overhead, 
leading to a Zero Overhead NoC (ZONoC) \cite{masera_noc}. 
The inter--processor communication needs are known a priori
as they depend on the structure of the ${\bf H}$ matrix. As a consequence, the best
path followed by a message during a decoding iteration can be statically derived for each
code and stored in form of
routing information distributed among nodes. This approach allows reducing packet size and complexity
of input FIFOs; moreover it eliminates the need for dynamic routing decisions
at NoC nodes. \\

As detailed in Section \ref{sec:configflow},
a dedicated cycle accurate simulation tool has been developed to configure the described NoC based architecture
for the decoding of a specific set of LDPC codes. This mainly implies deriving the
content of the routing memories and deciding the length of the FIFO memories. 
The tool receives a description of the NoC, the lists of parity check constraints that are mapped onto
each PE, and the scheduling of messages exchanged among NoC nodes. 
Using this information, the tool basically simulates the behaviour of the NoC while 
messages are injected by the PEs. At each node, incoming messages stored into input FIFO memories
are forwarded towards their destinations by means of a routing algorithm.
Routing decisions across a complete
decoding iteration are saved, together with the status of each FIFO. These decisions are then
translated into proper sequences of binary control signals
to be applied to FIFOs and crossbar switch: the ROUTING MEMORY (RM) indicated in Fig. \ref{fig:router}
is read at each cycle to provide router components with required control signals. The number of messages
stored at input FIFOs is continuously monitored to derive the required length for each FIFO.

\begin{figure}[t]
	\centering
		\vspace{-30pt}
		\hspace{-205pt}
		\includegraphics[scale=0.5]{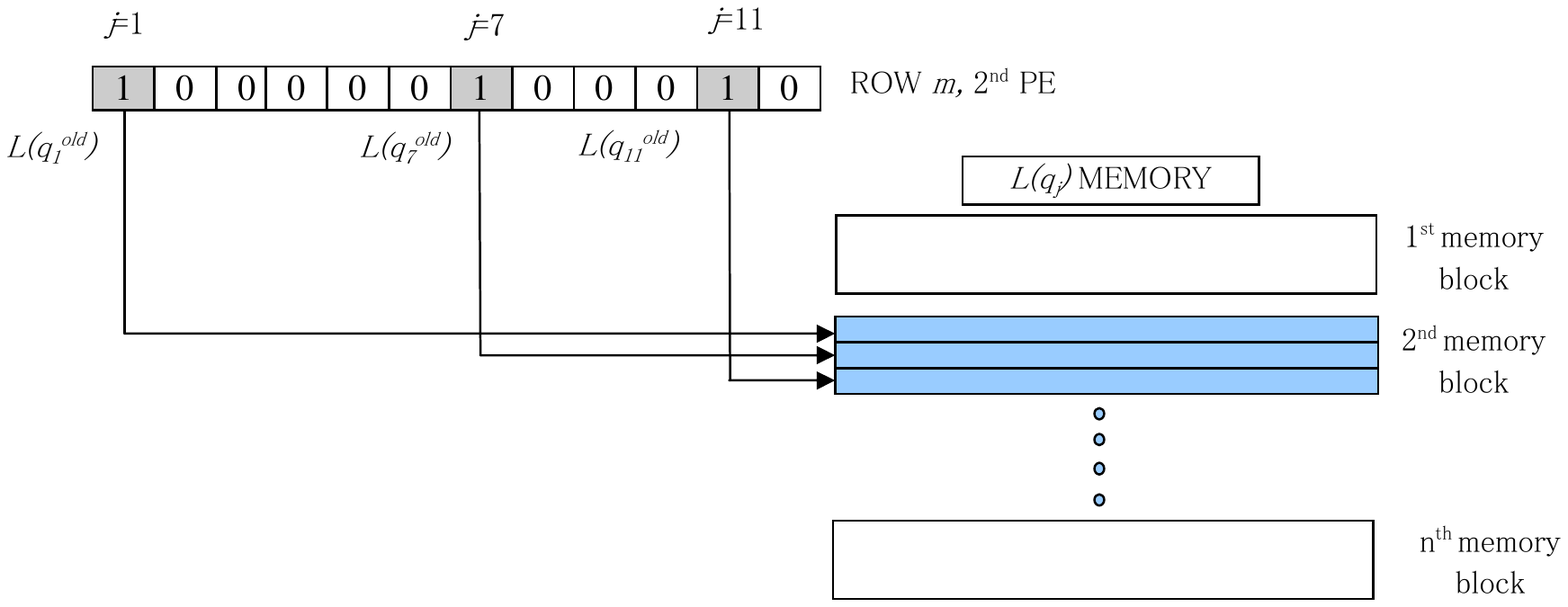}
		\hspace{-205pt}
		\vspace{-255pt}
	\centering
	\caption{Example of memory organization for extrinsic values}
		\vspace{-5pt}
	 \label{fig:esempio}
\end{figure}

\section{Architecture of the processing element}
\label{sec:architecture}

\begin{figure*}[t!]
  \centering
		\vspace{-50pt}
		\hspace{-130pt}
  \includegraphics[scale=0.65]{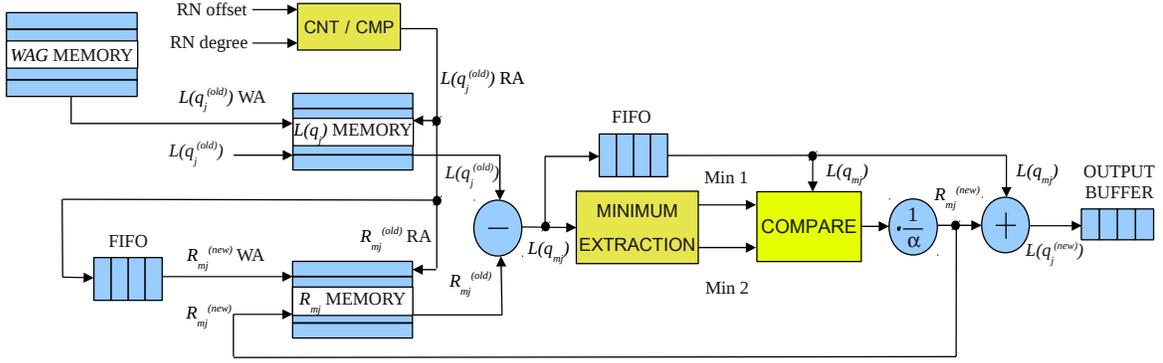}
		\vspace{-210pt}
		\hspace{-150pt}
		\caption{Simplified block scheme for the  processing element}
		\vspace{-5pt}
		\label{fig:decoder}
\end{figure*}

The general structure of the PE is shown in Fig. \ref{fig:decoder}.
Execution of equations (\ref{lq}) to (\ref{rmj1}) are organized in a pipelined way, in order to achieve high throughput.
Finite precision representation of data and number of decoding iterations have been decided by means of
extensive simulations of the considered LDPC codes.

Due to the use of an NoC as inter--processor communication structure, extrinsic information values 
necessary for the processing of a given parity check constraint are not received consecutively; instead a
PE receives extrinsic values related to multiple parity check constraints in an interleaved order.
This leads to the necessity of memories, to store received packets, and address generators to properly
retrieve stored packets. 
%


Extrinsic values $L(q_{j}^{(old)})$, generated at previous layer and sent through
the NoC, are received by the PE and stored in $L(q_{j})$ MEMORY. This two--port memory has $N_{pc} \times N_d$
locations, where $N_{pc}$ is the maximum number of parity check constraints mapped onto the PE and $N_d$
is the maximum degree of parity check constraints.

WAG MEMORY operates as write address generator. The sequence of extrinsic values received
at each PE is derived by means of off--line simulations and this information is used 
to initialize the WAG MEMORY with the list of addresses necessary to sequentially fill up $L(q_{j})$ MEMORY, while extrinsic values are received.
This means that all $L(q_{j}^{(old)})$ values required for a given parity check constraint 
are sequentially stored in the 
$L(q_{j})$ MEMORY, starting from an address equal to a multiple of $N_d$. Fig. \ref{fig:esempio} gives an example for
$L(q_{j})$ MEMORY organization. It is assumed that $N_{pc}$ parity constraints are mapped to the PE and
each parity constraint has $N_d=3$ degree. The memory is then divided into $N_{pc}$ blocks, each one containing 3 
consecutive locations. In the example, the 2$^{nd}$ scheduled parity check constraint receives 
three $L(q_{j}^{(old)})$
values from previous layer: these extrinsic values are sequentially stored in the 2$^{nd}$ block,
starting from offset $3$.

CNT/CMP component generates read addresses for $L(q_{j})$ MEMORY. 
As extrinsic values related to a given parity check constraint are sequentially stored, read addresses can be generated
by means of a counter. The counter is loaded with a proper offset to initially point to the location of
the first extrinsic value to be processed. 
The counter is then incremented to make accesses to the following values, up to a number
of read operations equal to $N_d$. A comparator is used to recognize the last read
operation and to load the counter with the offset required for the following parity check constraint to be served.

$R_{mj}$ MEMORY stores $R_{mj}$ amounts and has the same size as $L(q_{j})$ MEMORY.
The same address generators can be shared by the two memories; however, while read operations
are simultaneous for $L(q_{j})$ and $R_{mj}$ MEMORY, write operations to $R_{mj}$ MEMORY are delayed to accommodate 
the latency of the pipelined PE.

Updated $L(q_{mj})$ is derived from $L(q_{j}^{(old)})$ and $R_{mj}^{(old)}$ operands related to a previous layer.
These operands are read from respective memories and subtraction is performed.
First and second minimum values are then computed (MINIMUM EXTRACTION unit) and
the sign bits of all compared messages are XOR--ed.
The COMPARE unit implements (\ref{rmj1}) and its output is multiplied by $1/\alpha$ 
to obtain $R_{mj}^{(new)}$, which replaces
the previous value in $R_{mj}$ MEMORY.
%
Finally $L(q_{mj})$ is retrieved by means of a short FIFO and added to $R_{mj}^{(new)}$, so obtaining the new $L(q_{mj}^{(new)})$ (\ref{lq3}). 
The output buffer connects the PE to the NoC.


\section{NoC configuration}
\label{sec:configflow}

In order to support a specific LDPC code, the NoC and PE architectures introduced in the previous Sections 
must be configured.
This configuration includes three distinct operations: partitioning of parity check constraints over NoC nodes,
development of configuration data and uploading of configuration data to control memories.
In the first operation,  parity check constraints of every considered LDPC code 
must be clustered and each cluster must be assigned to one of the allocated PEs. 
In the second operation, the content of each control memory in both the NoC routing element 
and in the PE must be derived. 
In the third operation, control memories are updated with the derived configuration data to 
start working on a specific code.
The three configuration operations are detailed in the following sub--sections.

A preliminary need to design and configure the NoC based decoder is the decision on the 
required number of iterations and on the
finite precision representation to be adopted for the messages exchanged among PEs. These design choices 
heavily affect both decoding performance and implementation complexity, so they require careful consideration
and proper simulation models, described in
sub--section \ref{bersim}.

\subsection{Partitioning of parity check constraints}
Given the $\bf H$ matrix of the LDPC code, parity check constraints are partitioned and each partition
is assigned to a different PE. The main objectives of this 
mapping are the uniform distribution of the whole decoding effort among available PEs and the minimization of the
number of messages.
To achieve the first objective, approximately the same number of parity check constraints is mapped to each PE in the NoC. As for the second objective,
it is known that $\bf H$ matrices of LDPC codes do not show relevant adjacency among rows or columns, therefore the potential
advantages coming from optimal clustering of parity check constraints are limited. In this work, the structure of each code to be supported has been modeled as a graph, where each vertex corresponds to a parity check constraint and exchanged messages are associated to edges ($e_{ij}$ indicates the edge connecting vertexes $i$ and $j$).
The graph can be formally defined as $\mathbb{T}\mathcal{(V,E)}$ where $\mathcal{V}$ is the set of parity check constraints with cardinality $N_{pc}$, and the set of edges $\mathcal{E}$ is derived by listing all couples of
parity check constraints that share at least one bit, i.e. $\mathcal{E} = \{ e_{ij} |  i,j \in \mathcal{V},
\, N(i) \cap N(j) \not= \emptyset , \, i \not= j \}$.

The search for a good clustering of parity check constraints can be seen as 
a graph partitioning problem, which aims at solving the problem of dividing
$\mathbb{T}$ in $P$ partitions $\{\mathbb{T}_0(\mathcal{V}_0,\mathcal{E}_0), 
\mathbb{T}_1(\mathcal{V}_1,\mathcal{E}_1), \ldots , \mathbb{T}_{P-1}(\mathcal{V}_{P-1},\mathcal{E}_{P-1}) \}$, equalizing their size as much
as possible and trying to  minimize the \emph{cutset} cardinality. 
We define the \emph{cutset} $\mathcal{C}$ as a subset of
$\mathcal{E}$ that contains all those edges connecting two nodes
located in different partitions, i.e. 
\begin{displaymath}
\mathcal{C} = \{ e_{ij} \in \mathcal{E} | i \in \mathcal{V}_x, j \in \mathcal{V}_y, x \ne y, 0 \le x,y \le P-1 \}
\end{displaymath}
It is worth noting that each element of $\mathcal{C}$ represents a
message that has to be transmitted across the NoC. Hence
minimizing $\mathcal{C}$'s cardinality ($|\mathcal{C}|$)
corresponds to minimizing the overall number
of network flits exchanged over the network. \\
In this work a wrapper of the Metis graph partitioning library \cite{metis} 
(PyMetis Python package) has been used: the recursive k--way algorithm has been 
selected, where the number of partitions k is set equal to the number
of available PEs, $P$. The recursive application of the algorithm achieves a significant 
reduction in the number of exchanged messages
and this leads to a reduction of the global NoC traffic.
In Table \ref{tab:clustering}, a comparison in terms of exchanged
messages is shown for two partitioning techniques on several
different LDPC codes. Given every code, $\mathbf{|\mathcal{E}|}$ is the
total number of edges in $\mathbb{T}$, i.e. the number of messages that
nodes have to exchange every iteration. The two 
allocation strategies considered to map parity check constraints over the $P$ PEs of
the network are indicated as Random (RP) and Graph Partitioning (GP). 
In the Random strategy, parity check constraints have been randomly assigned to PEs,
under the constraint of uniform workload among nodes. Values reported in 
the RP column 
are actually the expected value of this RP process over 1000
realizations. Values in the GP column are obtained applying the recursive k--way algorithm 
to the $\mathbb{T}$ graph. It is clear that the GP approach leads to a
relevant saving of messages to be delivered with respect to the RP case.
This saving ranges between 34\% and 6\%, depending on the characteristics of the considered
LDPC code: in general, large saving percentages are obtained for low code rates and large
code sizes. The last column of Table \ref{tab:clustering} gives the   processing time required
to complete the GP on a 3 GHz Pentium 4 processor. This time strongly 
depends on the code characteristics, but it is fully affordable for all considered cases.

\begin{table*}[t!]
	\centering
	\caption{Effect of parity check constraints clustering on different LDPC codes and required processing times.
	$|\mathcal{C}|$ is the number of inter--PE messages; percentages are evaluated with respect to $|\mathcal{E}|$
	values.}
	\label{tab:clustering}
		\begin{tabular}{cccccc}
\hline
LDPC & $P$           & $|\mathcal{E}|$            & $|\mathcal{C}|$   & $|\mathcal{C}|$         & processing  \\		
code & number of PEs & initial number of messages & Random Partitioning (RP) & Graph Partitioning (GP) &times \\
\hline\hline
$802.16$e     &  25  & $7296$                     & $6908$            & $4800$  s               &$2.49$ s    \\
$(2304,1152)$ &      &                            & \emph{(94.6\%)}   & \emph{(65.8\%)}         &            \\

$802.16$e     &  25  & $7680$                     & $7370$            & $7061$  s               &$2.28$ s     \\
$(2304,384)$  &      &                            & \emph{(96.0\%)}   &  \emph{(91.9\%)}        &              \\

$802.16$e     &  25  & $5168$                     & $4935$            & $3412$   s              &$1.97$ s      \\
$(1632, 816)$ &      &                            & \emph{(95.5\%)}   &  \emph{(66.0\%)}        &              \\

$802.16$e     &  25  & $5440$                     & $5238$            & $4732$   s              &$1.60$ s       \\
$(1632, 272)$ &      &                            & \emph{96.3\%)}    &  \emph{(87.0\%)}        &              \\

$802.16$e     &  25  & $1824$                     & $1754$            & $1390$  s               &$1.11$ s       \\
$(576, 288)$  &      &                            & \emph{(96.1\%)}   &  \emph{(76.2\%)}        &              \\

$802.16$e     &  25  & $1920$                     & $1841$            & $1802$  s               &$1.14$ s      \\
$(576, 96)$   &      &                            & \emph{(95.9\%)}   & \emph{(93.8\%)}         &               \\

$802.11$n     &  16  & $6885$                     & $6428$            & $5387$ s                &$1.93$ s       \\
$(1944, 486)$ &      &                            & \emph{(93.4\%)}   & \emph{(78.2\%)}         &              \\

DVB-S$2$      &  64  & $71280$                    & $70180$           & $56516$                 & $ 159.91 $ s \\
$(16200, 6480)$ &    &                            & \emph{(98.5\%)}   & \emph{(79.3\%)}         &              \\

DVB-S$2$      &  64  & $285120$                   & $280598$          & $225409$                & $66.22 $ s      \\
$(64800, 25920)$ &   &                            & \emph{(98.4\%)}   &  \emph{79.1\%)}         &                  \\

Random code   &  25  & $3171$                     & $3042$            & $2644$                  & $2.41$ s       \\
$(1057, 244)$ &      &                            & \emph{(95.9\%)}   & \emph{83.4\%)}          &\\

\hline
		
		\end{tabular}
\end{table*}

\subsection{Development of configuration data}

The cycle accurate NoC simulator mentioned in Section \ref{sec:NoCdecode} must be run to
derive  configuration data for every LDPC code that has to be supported by the decoder. 
The simulator includes two main parts:
\begin{itemize}
\item a set of $P$ message generators, one for every PE, which inject messages into the NoC according to the decided
clustering
\item a complete model of the NoC, where $P$ routing elements receive messages from both neighbouring 
nodes and local PE and execute a routing algorithm to deliver them.
\end{itemize}
A message generator does not model the full PE, as the actual decoding is not required to configure the decoder; 
it simply scrolls a list of 
messages to be delivered and sends each of them through the NoC together with the identifiers related to source and 
destination nodes. Routing elements are modelled according to the structure shown in Figure \ref{fig:router}.
In the model, FIFO memories have a virtually unlimited length 
and the routing memory is replaced with a routing algorithm, which dynamically handles incoming messages. Different
routing algorithms can be adopted in the simulator, but the results given in this work have been obtained using
the very simple {\it O1Turn} algorithm proposed in \cite{01turn}. A complete simulation run comprises the injection
and delivery to final destinations of all messages exchanged among PEs on a single decoding iteration.
Since the same sequence of injection and routing operations is repeated for all decoding iterations, 
a single simulation run is enough for a specific LDPC code.
During a run, the simulator traces three kinds of data:
\begin{itemize}
\item the routing decisions made by the {\it O1Turn} algorithm at each NoC node and at each cycle 
\item the arrival order of messages at destination PEs
\item the number of occupied locations at each FIFO.
\end{itemize}
Routing decisions are converted into commands that must be stored in the routing memories (RM)
to properly control the hardware resources of  NoC nodes (input FIFOs, crossbar switch and 
output registers). Each memory has a length equal to the
number of cycles that are required to complete a whole decoding iteration.
The registered arrival order at each PE is used to fill the WAG memory. In addition
the CNT/CMP unit in Figure \ref{fig:decoder} must be initialized with identifiers of 
those parity check constraints that have been mapped
on the PE, and with the number of messages to be received. 
Finally the monitored numbers of occupied locations in the FIFO memories are used to decide on their length.

\subsection{Uploading of configuration data}

To prepare the NoC architecture for the decoding of a certain LDPC code, the generated configuration data must be uploaded 
to every NoC node. This process involves a considerable amount of data and can affect both occupied area and
throughput of the decoder. In wireless communications, adaptive coded modulation (ACM)
is a powerful technique capable to ensure maximum spectral efficiency while guaranteeing an acceptable BER level \cite{Viswanath}. 
In such techniques, the transmitter is allowed to switch between signal constellations and channel codes of varying size and rate at discrete 
time instants. Thus also the channel decoder must be able to dynamically switch between different codes. There are two options to implement this
switching: either the decoder is stopped during the configuration or concurrent data decoding and code reconfiguration are supported. 
The first solution has the drawback of reducing the decoder throughput and therefore it is only viable is the reconfiguration time is short and
the switch event unfrequent. Unfortunately this condition does not hold for most of standards.
As an example, in the WiMAX standard, information on channel condition is gathered 
by both base station and  subscriber station by averaging the feedback data received during 
a burst transmission \cite{wimaxstd}. 
When the computed average condition passes a certain threshold, a change of the current profile 
(code length, rate, modulation, frequency and power) is forced. 
The transmission of the new profile information spans over 
several OFDM symbols and takes a time in the order of milliseconds. As soon as the new profile is received, the decoder can 
reconfigure before the arrival of the new encoded frames. However, according to the standard, up to four different profiles can be 
clustered and sent together: in this case, the system is requested to switch between different profiles on a frame by frame base.
This implies that, during the decoding of the last received frame, the decoder must be reconfigured for the new code to be used
with the next frame. 

\begin{figure}[t]
	\centering
		\vspace{-60pt}
		\hspace{-110pt}
		\includegraphics[scale=0.5]{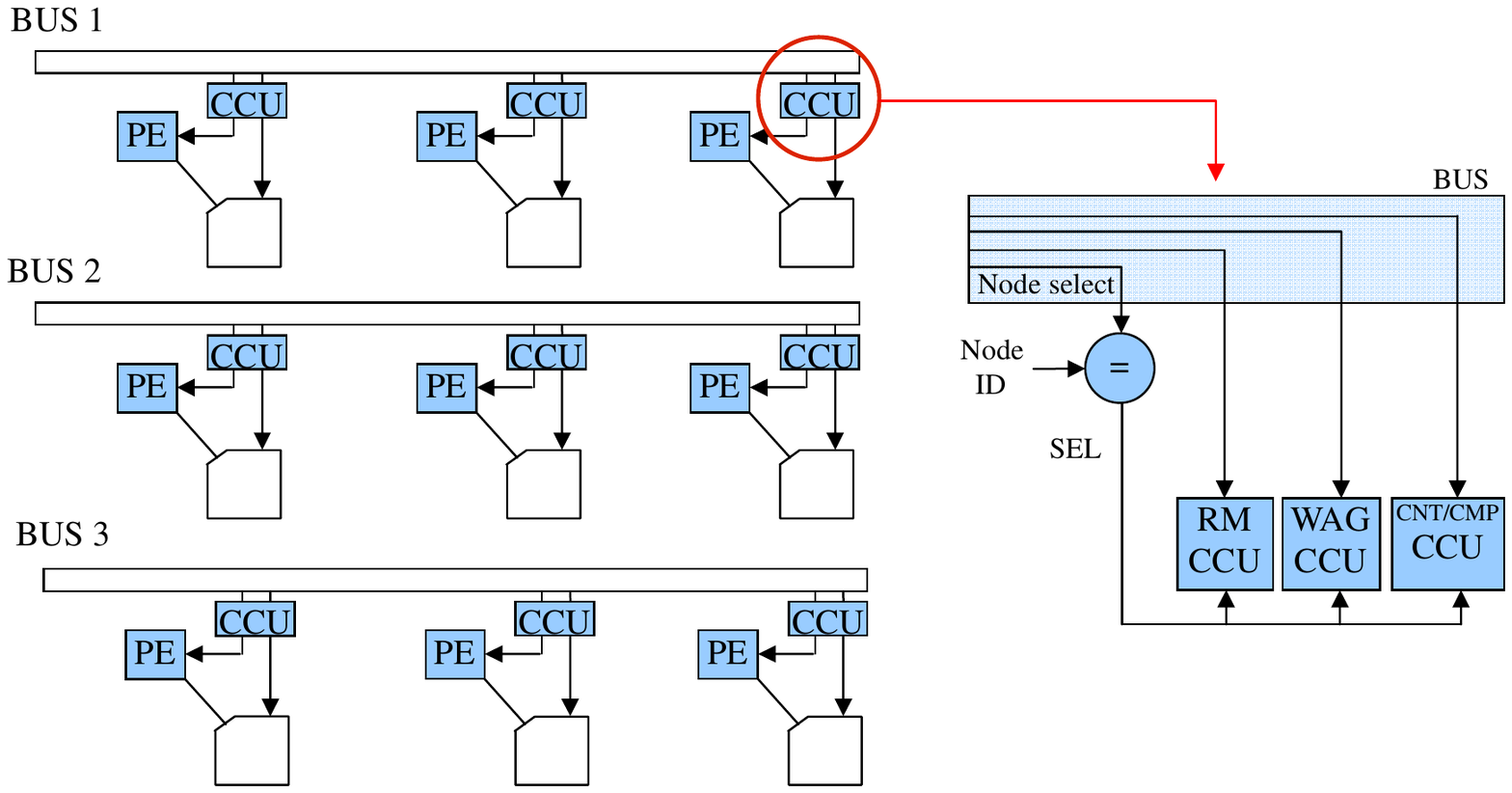}
		\hspace{-120pt}
		\vspace{-230pt}
	\centering
	\caption{Configuration of the NoC architecture}
			\vspace{-10pt}
	 \label{fig:config}
\end{figure}

In the $n \times n$ NoC based decoder, the upload is performed by means of $n$ parallel buses, one for each row of the NoC (Figure \ref{fig:config}). 
Every bus sequentially updates the $n$  nodes of a row. For every node, the WAG, RM and CNT/CMP components must be
written; each of these memories needs to receive a number of words equal to the number of clock cycles 
required for a single decoding iteration. This number changes from code to code: let us 
indicate it as $k_i$ for code $C_i$.
Each bus is composed of enough parallel lines to simultaneously carry one configuration word for each 
WAG, RM and CNT/CMP component, plus an identifier for the target node. Thus $n \times k_i$ cycles are required to complete the uploading of an $n \times n$ architecture with the configuration data related to code $C_i$.
For the case of a $5 \times 5$ NoC decoder used to operate on WiMAX codes (Section \ref{sec:results}), 
10 bits are necessary for the WAG memory, 15 for the RM memory and 3 bits for the node identifier. 
Overall $38$ lines per bus are needed. 
As shown in Figure \ref{fig:config}, a Configuration Control Unit (CCU) connects bus lines to 
node components that must be programmed: in the CCU, the node identifier (Node ID) is used to select the addressed node (SEL signal);  
the other fields in the bus are extracted and delivered to their destinations, which are three local Configuration Control Unit:
one for RM in the routing element, and two for WAG and CNT/CMP in the PE. 

In order to enable concurrent decoding and uploading of new configuration data, RM, WAG and CNT/CMP components
are organized as circular buffers. 
Let us indicate as $B$ the capacity of the buffer;
$K_{max}$ is the maximum value of $k_i$ over the set of codes to be supported. 
When switching from code $C_1$ to code $C_2$, the current configuration data ($k_1$ words) must be discarded 
and replaced with the new ones ($k_2$ words).
If $B=2 k_{max}$, then the circular buffers are long enough to contain at the same time the 
configuration data for both $C_1$ and $C_2$, even in the worst case $k_1=k_2=k_{max}$. 
With this choice of $B$, the uploading of $C_2$ configuration data
can be distributed along multiple decoding iterations performend on $C_1$ code and the decoder
can switch to $C_2$ as soon as required, without pausing the decoding activity.
Double port memories are required to implement such circular buffers.

However the length of the circular buffers can be reduced by splitting the uploading process into the following three phases: 
\begin{enumerate}
\item partial uploading of the new configuration data into the section of the circular buffers that is not used for $C_1$
\item partial uploading during the last decoding iteration performed on $C_1$
\item partial uploading during the first decoding iteration performed on $C_2$.
\end{enumerate}
As current code $C_1$ uses $k_1$ locations of every circular buffer, a number $B-k_1$ of configuration words can be uploaded
without overwriting useful information: therefore phase 1 can be started as soon as the code switching has been decided and 
it can be distributed along one or multiple decoding iterations. If $k_2 < B-k_1$, phases 2 and 3 are not necessary,
otherwise additional $k_1/n$ configuration data can be uploaded during phase 2: $k_1$ cycles are available in phase 2, but 
the $n$ nodes connected on a same bus must be uploaded sequentially, thus only $k_1/n$ cycles can be used for each node in a row. 
The partial uploading of phases 1 and 2 is able to write $B-\frac{n-1}{n}k_1$ words to the configuration memories. 
If $k_2 > B-\frac{n-1}{n}k_1$, also phase 3
can be exploited: in this case, the uploading is completed while the already written words of the new
configuration are read to control the decoding on $C_2$. Phase 3 provides additional $k_2/n$ cycles. 
Therefore the overall number of cycles available in the three phases 
to load all $k_2$ configuration data is $t=B-\frac{n-1}{n}k_1+ \frac{k_2}{n}$. The condition $t>k_2$  leads to 
\begin{equation}
	\label{buffersize}
     B > \frac{n-1}{n} \left( k_1 + k_2 \right)
\end{equation}
In the worst case ($k_1=k_2=k_{max}$), $B$ must be equal at least to $2 \frac{n-1}{n} k_{max}$. 
For example, if $n=5$ 
($5 \times 5$ NoC decoder), $B > 1.6 k_{max}$, which corresponds to 20\% of saved area with respect to the initial
assumption $B=2 \cdot k_{max}$. 
In WiMAX, the worst case is obtained when $C_1$ and $C_2$ are the codes with rate $0.75$ and length $2208$
and $2304$ respectively. In this case, $k_1=k_{max}=491$ cycles and $k_{2}=466$ cycles.
Working with a $5\times5$ NoC and using $n=5$ buses, the required size for the circular buffers is $B=767= 1.56 k_{max}$.

\begin{figure}[t]
	\centering
		\vspace{-60pt}
		\hspace{-110pt}
		\includegraphics[scale=0.5]{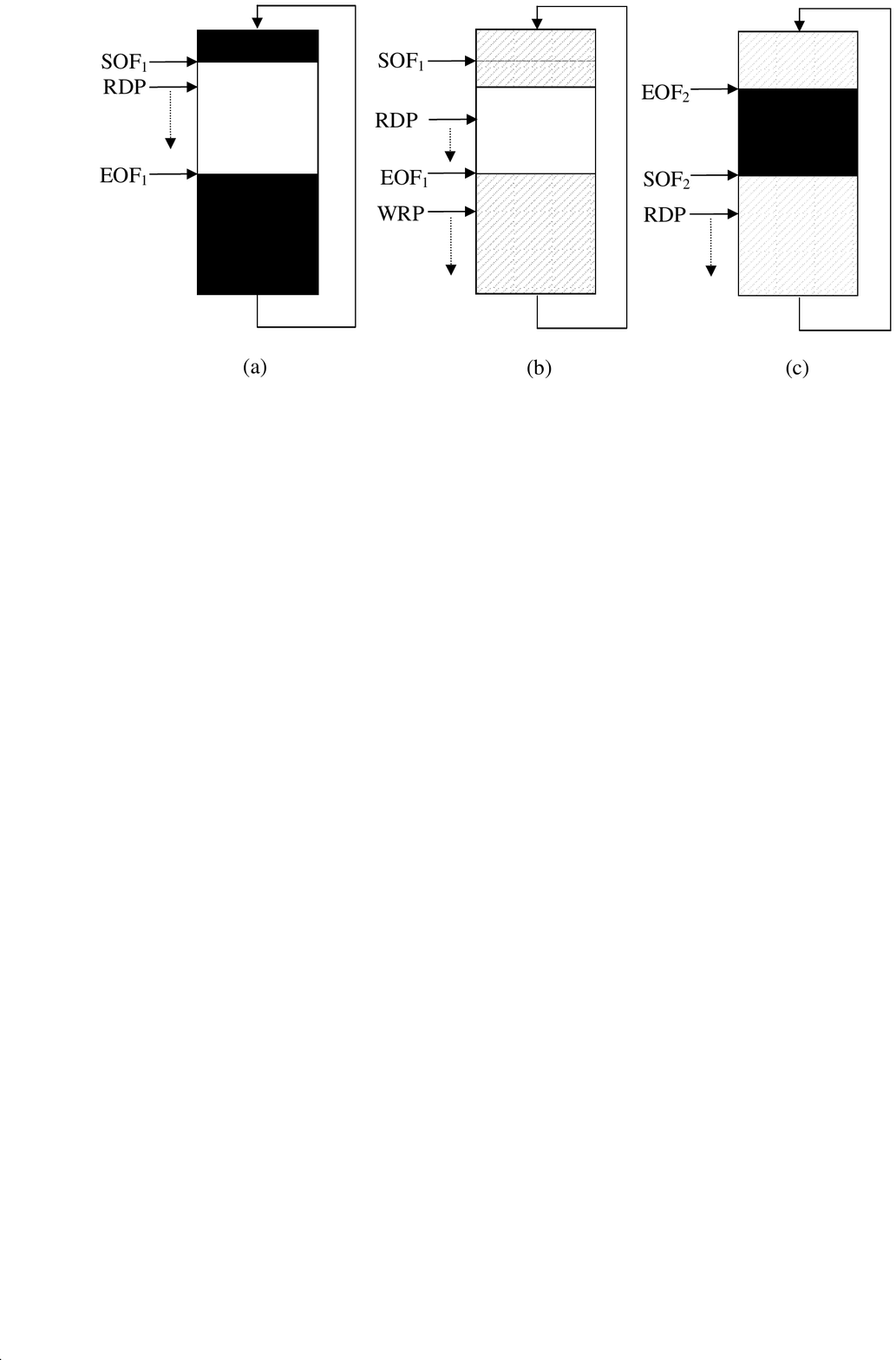}
		\hspace{-120pt}
		\vspace{-270pt}
	\centering
	\caption{Handling of configuration memories during upload of a new code: unused (black), $C_1$ (white), $C_2$ (grey striped)}
			\vspace{-5pt}
	 \label{Fig:reconfig_mem}
\end{figure}

In Figure \ref{Fig:reconfig_mem} various steps of the upload process are presented. 
Figure \ref{Fig:reconfig_mem}.(a) represents the circular buffer status during the decoding on $C_1$, 
while no reconfiguration is active. 
Configuration words for $C_1$ are stored in $k_1$ consecutive locations (white area) between limits contained in registers
Start of Frame ($SOF_1$) and End of Frame ($EOF_1$). 
The Read Pointer ($RDP$) is used by the decoder to access circular buffers during the decoding process.
In configuration phase 1 (Figure \ref{Fig:reconfig_mem}.(b)), write to circular buffers 
is handled by means of Write Pointer ($WRP$), which is initialized at $EOF_1+1$ and incremented at each new write
(grey striped area). 
Phase 1 is stopped when $WRP$ reaches $SOF_1$. Uploading of the circular buffer is started again
at the beginning of the last planned iteration on code $C_1$ (phase 2): both $WRP$ and $RDP$ are increased 
between $SOF_1$ and $EOF_1$ in this phase, to replace old configuration words with new ones.
At phase 3, $SOF_1$ and $EOF_1$ are updated according to the characteristics of
code $C_2$: $SOF_2=EOF_1+1$ and $EOF_2 = (SOF_2 + k_2)mod B$ (Figure \ref{Fig:reconfig_mem}.(c)). 
In this phase, while $RDP$ 
is incremented to read the first words on the new configuration, remaining words located before 
$EOF_1$ are written to complete the uploading. 

\subsection{Simulation of LDPC decoding}
\label{bersim}

A C++/Python finite precision model has been developed to simulate LDPC codes. 
Several choices can be made when running the model, such as
decoding algorithm (e.g. Sum--Product, Min--Sum, normalized Min-Sum), 
scheduling (two--phase and layered), floating or fixed point representation of data. 
In addition the maximum number of iterations $It_{max}$ can be programmed and different methods for early stopping of the decoding can be supported.
This model is used with the 
purpose of driving some design choices, which affect bit error rate (BER) performance, 
implementation complexity and throughput:
\begin{enumerate}
\item the appropriate number of iterations for each specific code 
\item the optimal value for the \emph{$\alpha$} parameter in the normalized Min--Sum algorithm
\item the proper representation for extrinsic information and any other amount processed by the decoding algorithm.
\end{enumerate}
The first two choices, the number of iterations and the value of $\alpha$, can be adapted to
each specific code to be supported, while a unique data representation must be decided for
all codes.

\begin{figure}[t!]
  \centering
		\vspace{-15pt}
		\hspace{-170pt}
  \includegraphics[angle=270,scale=0.37]{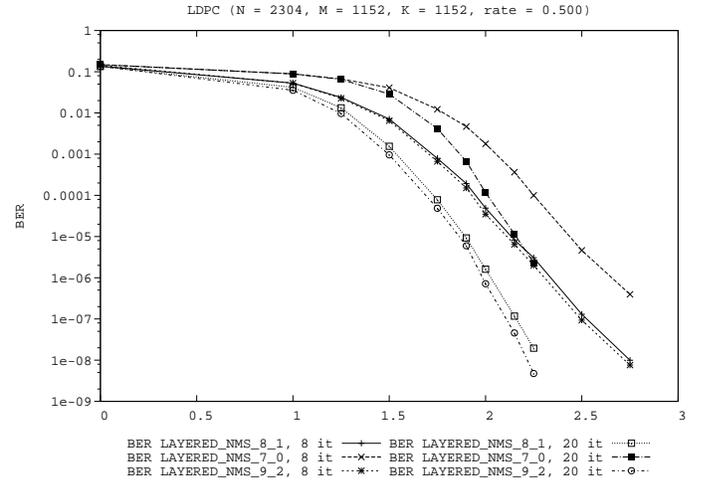}
		\vspace{-10pt}
		\hspace{-150pt}
		\caption{BER curves for 9, 8 and 7 quantization bits and for 8 and 20 maximum iterations}
		\vspace{-10pt}
		\label{fig:ber}
\end{figure}

\begin{figure}[t!]
  \centering
		\vspace{-15pt}
		\hspace{-170pt}
  \includegraphics[angle=270,scale=0.37]{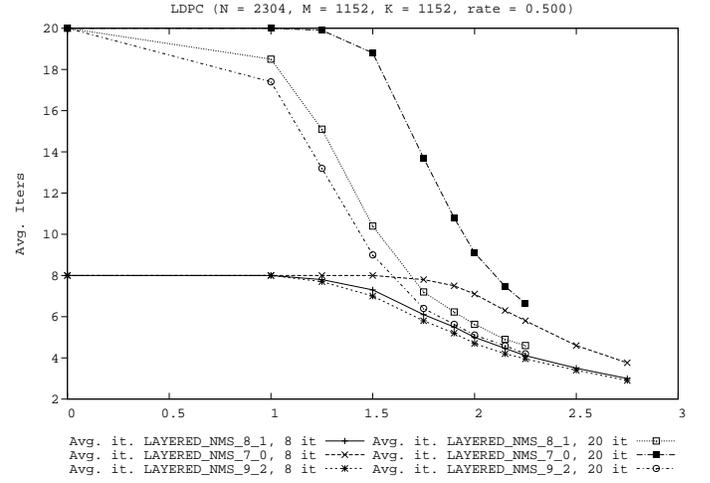}
		\vspace{-10pt}
		\hspace{-150pt}
		\caption{Average number of iterations.}
		\vspace{-10pt}
		\label{fig:iters}
\end{figure}

As an example, in Fig. \ref{fig:ber}, BER curves obtained via the C++/Python model are 
reported for the $2304\times1152$, 
rate $0.5$  WiMAX code, which is one of the cases of study described in the following Section;
the average number of completed iterations is plotted in Fig. \ref{fig:iters} for the same code. 
The curves refer to the layered normalized Min--Sum algorithm, with  $It_{max}$ equal
to either 8 or 20 and multiple choices of message quantization.
The notation $n\_m$ used in the legend of Fig. \ref{fig:ber} and \ref{fig:iters} 
corresponds to the allocation of a total of $n$ bits to
represent extrinsic values, with $m$ bits used to indicate the fractional part: it can be seen that
9\_2 and 8\_1 choices offer almost the same performance. From the high level model 
the normalization factor $\alpha$  has been set to 1.15
for the WiMAX codes.

\begin{table*}[t!]
	\centering
	\caption{LDPC architectures comparison: CMOS technology process (TP), area occupation (A), normalized area occupation for 65nm technology (An), clock frequency ($f_{clk}$), precision (P), maximum number of iterations ($It_{max}$), throughput ($T$), average throughput ($T(av)$), and SNR to achieve BER=$10^{-5}$ ($SNR$)}
	\label{tab:compare}
		\begin{tabular}{|c|c|c|c|c|c|c|c|c|c|c|c|c|}
		\hline
		Decoder                   & Tp       & A     &An    & $f_{clk}$         & P          &$It_{max}$     &It   &Code & $T$  & $T (av) $   & SNR  & Flexibilty \\
		                                 & [nm]   & [mm$^2$] & [mm$^2$]  & [MHz]   &[bits]  	&  	   &(average)   & length - rate   & [Mb/s]  & [Mb/s]  & [dB] & \\
		\hline	
		
		\cite{flexichap} 	 & 65         &0.23   & 0.23    &  400       & N/A     &  20  &  N/A  &WiMAX   	&  27.7   & N/A &N/A  &YES\\	
	\cite{Liu_2009}           	 & 90        & 6.22    &3.24   &  300        & 6     &  20  & N/A 	 &WiMAX 	&  212 (max)  & N/A  &2.2  (min)  &YES\\
		\cite{flex_Huo}   	 & 180        & 3.39   &0.442    &  100       & N/A     &  10  &  N/A  &WiMAX   	&  68   & N/A &N/A  &YES\\	
\cite{Chich}           	 & 90        & 6.25     &3.26    &  109       & 6     &  20  & N/A 	 &WiMAX &  63   & N/A &2.2  (min)  &PARTIAL\\
	\cite{Xin}               	 & 130      &  8.29    &2.075    &  83         & 8     &  8    & N/A &WiMAX   	&  60 (min)  & N/A  &3.1  (min)  &YES\\	
\cite{Brack}           	 & 65        & 1.337   &1.337    &  400       & 6     &  20  & N/A  &WiMAX   	&  48 (min)  & N/A &N/A  &PARTIAL\\
\cite{flex_Huang}       	 & 130        & 6.3    &1.575   &  260       & 4     &  15   & N/A &WiMAX     & 205 (max)  & N/A &2.15  (min)   &YES\\	

		\cite{masera_noc} 	 &130       & 3.7      &0.93     &  300       & 6     &  10   & N/A	 &$2304$ - $0.5$   &  56  & N/A  & N/A  &YES\\

		\cite{Yeong}-proposed & 180      & N/A    & N/A    &   200       & 9     &  N/A  &  4.6 &$2304$ -  $0.5$   & N/A  	&  106   &1.9  &NO\\
		\cite{Yeong}-layered    & 180      & N/A     & N/A   &   200       & 9     &  N/A  &  5.6 &$2304$ -   $0.5$  & N/A  	&  71   &1.9  &NO\\
		\hline
		\cite{Yu}           	 & 90        & 0.679   &0.354    &  400       & 7     &  
\begin{minipage}[c]{1cm}
\centering
\vspace{0.5mm}
12\\
8
\vspace{0.5mm}
\end{minipage}   &  
\begin{minipage}[c]{1cm}
\centering
\vspace{0.5mm}
6.64\\
3.66
\vspace{0.5mm}
\end{minipage} & 
\begin{minipage}[c]{1.5cm}
\centering
\vspace{0.5mm}
$2304$ -  $0.5$\\
$2304$ -  $0.83$
\vspace{0.5mm}
\end{minipage}   & 
\begin{minipage}[c]{1.2cm}
\centering
\vspace{0.5mm}
66.7\\
200
\vspace{0.5mm}
\end{minipage} &  
\begin{minipage}[c]{1cm}
\centering
\vspace{0.5mm}
N/A \\
N/A 
\vspace{0.5mm}
\end{minipage}  &
\begin{minipage}[c]{1cm}
\centering
\vspace{0.5mm}
2.15\\
3.95 
\vspace{0.5mm}
\end{minipage} &
PARTIAL \\
		\hline
		\hline	
	&      &     &     &         &     & 10    &2.9      &$576$ -  $0.5$	&  71 (min.) & 244 	  &2.9   &\\
&      &     &     &         &    & 14   &  1.9    &$576$ -  $0.83$	&  80 (min.) & 592	  & 4.3  &\\
&      &     &     &         &    & 10    & 4.9     &$1632$ -  $0.5$	&  78 (min.) & 159 	  & 2.4  &\\
{\bf $5\times5$ NoC}	& 130      & 4.72    &1.18    &  300       & 8  & 14    &  2.8    &$1632$ -  $0.83$	&  84 (min.) & 588	  & 3.9  &YES\\
(WiMAX) &      &     &     &         &    & 10    &  6.1    &$2304$ -  $0.5$	&  82 (min.) & 135 	  & 2.2  &\\
&      &     &     &         &     & 14    & 2.7     &$2304$ -  $0.83$	&  89 (min.) & 462	  & 3.7  &\\

\hline
\hline
\begin{minipage}[c]{2cm}
\centering
\vspace{0.5mm}
{\bf $4\times4$ NoC}\\
(WiFi)
\vspace{0.5mm}
\end{minipage}	& 130      & 3.35    &0.838     &  300       & 8 & 15    &  3.8    &$1944$ -  $0.75$	&  73 (min.) & 288 	  & 3.0  &YES\\
\hline
\hline

\begin{minipage}[c]{2cm}
\vspace{0.5mm}
\centering
{\bf $8\times8$ NoC}\\
(DVB-S2)
\vspace{0.5mm}
\end{minipage} & 130      & 13.98    &3.494     &  300       & 8 &  
\begin{minipage}[c]{1cm}
\centering
\vspace{0.5mm}
12\\
12
\vspace{0.5mm}
\end{minipage}   &  
\begin{minipage}[c]{1cm}
\centering
\vspace{0.5mm}
4.3\\
4.5
\vspace{0.5mm}
\end{minipage} & 
\begin{minipage}[c]{1.4cm}
\centering
\vspace{0.5mm}
$16K$ - 0.6\\
$64K$ - 0.6
\vspace{0.5mm}
\end{minipage}   & 
\begin{minipage}[c]{1.2cm}
\centering
\vspace{0.5mm}
90 (min.)\\
92 (min.)
\vspace{0.5mm}
\end{minipage} &  
\begin{minipage}[c]{1cm}
\centering
\vspace{0.5mm}
188\\
195 
\vspace{0.5mm}
\end{minipage}  &
\begin{minipage}[c]{1cm}
\centering
\vspace{0.5mm}
2.25\\
2.1 
\vspace{0.5mm}
\end{minipage} & YES\\
		\hline

		\end{tabular}
\end{table*}

\section{Cases of study and achieved results}
\label{sec:results}

In order to show the feasibility of the described approach to flexible LDPC decoding, a $5 \times 5$ NoC based 
decoder has been sized to support the whole set of WiMAX LDPC codes and designed for a 130 nm standard cell technology.
Main design choices for this set of LDPC codes are:
\begin{itemize}
\item representation of extrinsics, according to notation in Sub--section \ref{bersim}, 8\_1 
(sufficient to guarantee good BER performance)
\item normalization factor, $\alpha=1.15$
\item maximum number of iterations, 10 and 14
\item length of the FIFOs, 7 (obtained through NoC simulator)
\end{itemize}

The decoder architecture relies on a 25--PEs torus mesh, and its sizing refers to the
largest code in the WiMAX standard: code block size $2304$ and $N_d = 20$. 
The number of PEs has been selected to guarantee a throughput of at least 70 Mbits/s.
The whole decoder has been described using VHDL language and synthesized 
with Synopsys Design Compiler. 
The results in terms of 
area occupation and throughput for the proposed case of study are shown in the second part of Table \ref{tab:compare} 
(rows related to the $5 \times 5$ NoC based decoder tuned for WiMAX codes).
The WiMAX standard includes several combinations of code lengths and code rates. Among them, 
three lengths ($57$6, $1632$ and $2304$) and two
code rates ($0.5$ and $0.83$) have been reported in the Table to show the throughput offered by the proposed decoder.\\
Several other implementations supporting the same standard are provided in the first part 
of Table \ref{tab:compare} to enable comparisons. It can be seen that, notwithstanding its large flexibility, 
the proposed solution is compliant with the throughput requirements imposed by the standard.
Moreover the achieved results show better worst-case throughput than all the compared decoders, while the 
provided average throughput value  shows the gain that can be achieved with the introduction of an 
early iteration stopping mechanism \cite{chen2010}.
It is worth noticing that the area complexity provided for the NoC based decoder also
includes the overhead deriving from the reconfiguration procedure described in
Section \ref{sec:configflow}: this overhead corresponds to 14.5 for the $5 \times 5$ NoC.

To show that the proposed decoder is not limited to the WiMAX standard, but can actually work with every LDPC
code with smaller length, the results for a random code of size $1057$ and rate $0.77$ are also provided in Table \ref{tab:compare}
(row labelled as ``$5 \times 5$ NoC, random code''). This code is fully supported by the same decoder designed for the
WiMAX codes: it achieves a throughput of 147 Mbit/s with $It_{max}=8$ and a BER$=10^{-5}$ at SNR$=2.3$ dB.\\
The other implementations included in Table \ref{tab:compare} exhibit a flexibility limited to the set
of WiMAX codes and cannot work on different codes: particularly they do not support codes with a random structure
of the parity check matrix. Moreover most of them use less than 8 quantization bits to represent extrinsics.
Notwithstanding these differences, the overall 
area occupation of the proposed decoder (An, nomalized area at 65 nm) is lower than most of the compared solutions.

A second NoC based decoder has been designed and sized in order to comply with the IEEE $802.11$n WiFi standard. 
Qauntization and normalization factor are the same as for the WiMAX case, while FIFO length has been 
reduced to 3 and $It_{max}=15$.
Since the  involved codes are smaller than in WiMAX, a $ 4 \times 4$ NoC with 16 PEs is adopted. 
The area occupation is greatly reduced with respect to the $5\times5$ solution, while the reconfiguration overhead remains almost the same (15.1\%). The BER crossing point is fairly low, and the low number of average iterations 
leaves room for considerable throughput improvements in case of presence of an early stopping criterion. 

Finally, a third NoC based decoder has been sized to support the DVB-S2 standard. The 
block length for this code is much larger 
than for WiFi and WiMAX standards and an $8 \times 8$ NoC is necessary to obtain a sufficient throughput ($90$ Mb/s). 
FIFO maximum depth is equal to 15 in this case and $It_{max}=12$.\\
The occupied area is very large for this 64 PEs decoder, with a  12\% overhead due to reconfiguration circuits. 
However
the full flexibility offered by the NoC approach makes it possible to map on the same architecture
any of the decoders described above for WiMAX or WiFi codes.

\section{Conclusions}
\label{sec:conclusions}

The design of a NoC based LDPC decoder is presented. The decoding architecture is fully flexible
in terms of supported codes and adopts normalized min--sum algorithm with layered scheduling. 

Current flexible solutions manage to reach only partial flexibility by concentrating on a subset of codes; major modifications 
are required in these decoders to extend the kind of supported codes. Previous NoC based solutions, on the other side, offer larger
flexibility but fail to provide acceptable throughput. In the results section it is shown that the proposed decoder 
can guarantee very good performance and full flexibility at the cost of a small increase of area occupation w.r.t.
dedicated decoders. Moreover the decoder takes advantage of NoC scaling properties, enabling the allocation of different numbers of PEs, according to the desired area throughput trade--off.

\bibliographystyle{IEEEtran}
\bibliography{IEEEabrv,IEEEexample}

\end{document}